\documentclass[12pt]{article}

\usepackage{amsmath}
\usepackage{graphics,graphicx,epsfig}
\usepackage{amssymb}
\usepackage{epstopdf}
\usepackage{color}

\usepackage{sidecap}
\usepackage{appendix}

\def\L3{ {\Lambda \over 3}}

\def\obsone{{\cal O}_1}
\def\obstwo{{\cal O}_2}
\def\lt{\tilde{L}}

\def\be#1\ee{\begin{equation}#1\end{equation}}
\def\bea#1\eea{\begin{align}#1\end{align}}

\begin{document}

\begin{titlepage}
\vfill

\vfill
\begin{center}
\baselineskip=16pt
{\Large\bf Phase Space of SdS Geodesics and using the }\\
 {\Large\bf  Cosmological Horizon to Observe a Black Hole }

\vskip 10.mm

{\bf   Bela Nelson $^{(a)}$, Allison Powell$^{(b), (c)}$,   and Jennie Traschen$^{(b)}$ } 

\vskip 0.4cm

{$^{(a)}$ Department of Mathematics, University of Massachusetts, Amherst, MA 01003, USA}\\
{$^{(b)}$ Department of Physics, University of Massachusetts, Amherst, MA 01003, USA}\\
{$^{(c)}$ Department of Physics, Brandeis University, Waltham, MA 02453, USA}\\
\vspace{0.3cm}
\vskip 4pt

\vskip 0.1 in Email: \texttt{binelson@umass.edu, acpowell@brandeis.edu, traschen@umass.edu}
\vspace{6pt}
\end{center}
\vskip 0.3in
\par
\begin{center}
{\bf Abstract}  
\end{center}
\begin{quote}
Light propagating from near a black hole horizon to the outside world is highly redshifted. In the limit that the emitter passes through 
the horizon, the redshift becomes infinite. In this sense the near horizon region is unobservable, as emission energies fall below some detectability bound.
However, in Schwarzschild de Sitter (SdS) spacetime there is a second, cosmological, horizon due to the positive cosmological constant. Judiciously placed
observers can take advantage of the blueshift due to this horizon.  The frequency of signals emitted from near the black hole can be shifted back
 upward to an observable 
value. This effect is computed for a variety of accelerated and geodesic observers. An analysis of radial and circular geodesics in SdS is a key 
component of the paper.
We find a  ``cresting-wave" shaped critical curve in the SdS-geodesic parameter space such that under the curve there are three circular orbits, on
the curve there are two orbits, and elsewhere
there is one. It is found that the best strategy for observing photons from an emitter falling into the black hole is for the receiver 
to be near the cosmological horizon and
also moving towards the black hole. For photons emitted from the smallest circular orbit and received at the largest  circular orbit,
the nonzero cosmological constant enhances observations by a factor that varies from zero to three.

\vfill
% \hrule width 5.cm
\vskip 2.mm
\end{quote}
\end{titlepage}

\section{Introduction}
Students and teachers of general relativity, from time to time, run into the following question:
how is it possible for us to observe a black hole when, from the perspective of an outside observer, we can never observe anything falling into the black hole? The
question refers to the fact that light emitted from an infalling device is redshifted as it propagates out of the gravitational well created by
a black hole, and in the limit that the device crosses the horizon the redshift becomes infinite. 
From the perspective of someone falling into the black hole, the person simply falls through in finite proper time,
 but from an outside perspective, it takes an infinite amount of time for the person to fall in.
 
 Schwarzschild-de Sitter (SdS) spacetime describes a black hole with mass $M$ in a universe with a positive cosmological constant $\Lambda$. 
 The black hole horizon is surrounded by a second, cosmological, horizon. From the point of view of an observer ${\cal O}$ in the region between
 the two horizons, the black hole pulls things in and the cosmological expansion pulls things out. Light emitted from near either
 horizon and received by a generic ${\cal O}$ is heavily redshifted. On the other hand, if ${\cal O}$ is near the cosmological horizon, though
a photon that is emitted from near the black hole is first redshifted it then experiences a compensating blueshift as it propagates towards the cosmological
 horizon. Hence the energy of signals can be shifted back into a detectable range for ${\cal O}$.
 
 The simplest situation is when the emitter $\obsone$  and the receiver $\obstwo$ are static,
with $\obsone$ near the black hole horizon and $\obstwo$ near the cosmological horizon.
 To maintain their position each must be accelerating. The ratio of the received to emitted energies is found to be
 \be\label{introratio}
 {E_{2 } \over E_{1 } } \simeq {\kappa_b^{3/2} \over | \kappa_c |^{3/2} } {|a_2 | \over |a_1 |}
 \ee
 where $\kappa_h , \ h=b,c$ are the horizon surface gravities and $|a_I | , \ I=1,2$ are the magnitudes of the accelerations of the two observers. 
 So in our thought experiment, the magnitudes of the accelerations can be adjusted to yield a ratio of energies that is detectable.
 This paper explores the redshift/blueshift effect for various choices of static and then geodesic observers.

 Geodesic observers are more natural than accelerating ones since no extra force is required. So we proceed by first studying 
 geodesics of SdS and what the possible orbits are in the $M, \Lambda , L$ parameter space, where $L$ is the angular momentum of
 the particle. We find  a curve of critical points in the parameter space
 which has the shape of a cresting wave (see Figure (\ref{fig:gamma,alpha}))  on which there are two orbits. 
 Inside the curve there are three circular orbits, and  outside the curve there is one.
In addition, SdS has an unstable
  static, zero angular momentum geodesic, 
 which illustrates the balance between the pull of the black hole and the cosmological expansion. We show that the larger unstable orbits are
 perturbatively connected to a circle of static solutions as the angular momentum of the particle goes to zero.
 Hence the qualitative features of these simple geodesics in SdS are different from Schwarzschild in interesting ways.
 
 Lastly we compute the redshifts for various choices of geodesic emitters and receivers. It turns out that
 the best strategy for observing the near horizon region of
 the black hole via signals sent by $\obsone$
  is to arrange for observer $\obstwo$ to be near the cosmological horizon and falling towards the black hole. Of course as $\obstwo$ gets
  farther from the cosmological horizon the signals get more redshifted. This also works if $\obstwo$ is outside, but near, the cosmological
 horizon, who again benefits by a blueshift. More leisurely measurements can be made by observers in circular 
 orbits, though there are no orbits arbitrarily close to either horizon. Still, there is an enhancement factor in this case, compared to $\Lambda =0$,
 which we compute.
 
   A full integration of the geodesics in SdS and Kerr-SdS can be found in \cite{Hackmann:2010zz}\cite{Lammerzahl:2015qps}.
The form of the results is not easy to work with for our 
 applications, so in this paper we use analytic tools to get more limited, but simpler, results which are sufficient to illustrate the redshift/blueshift physics. Reference \cite{Momennia:2023lau}
 contains a detailed analysis of rotating black holes in de Sitter, working in the limit of small $\Lambda$ to get results appropriate for observations of
 black holes in the current epoch of our universe. Images and plots constructed from numerical integrations of the geodesics in SdS are presented in references
 \cite{Wang:2023fge} \cite{Cao:2024kht}, and show enhancement  of the received luminosity for observers close to the cosmological horizon.

 This paper is organized as follows. Section 2 summarizes basic information about the geometry of SdS, and Section 3 computes the redshift/blueshift 
 effect for static observers. Section 4 analyzes radial and circular geodesics, and Section 5
 applies the results to computing the redshift for various choices of observers. For readers uninterested in the  derivations about geodesics, Section 4 can be 
 skipped without loss of continuity. Section 6 extends the analysis to observers beyond the cosmological horizon and Section 7 concludes the paper.
 
\section{Schwarzschild de Sitter}
The Schwarzschild deSitter metric describes a black hole in a spacetime with positive cosmological constant $\Lambda$. The metric is
 \begin{equation}\label{metric}
    ds^2=-f(r)dt^2+\frac{dr^2}{f(r)}+r^2d\Omega^2,
\end{equation} with
 \begin{equation}\label{fdef}
    f(r) = 1-\frac{2M}{r}-\frac{\Lambda}{3}r^2 
\end{equation}
The horizons occur at the positive zeros of $f(r)$, so it is also useful to write $f$ in the factored form
\begin{equation}\label{fdef2}
f(r) = -{\Lambda\over 3r}(r-r_c)(r-r_b)(r-r_n)
\end{equation}
We set $r_c \ge r_b$.
Comparing the two forms of $f$ gives $r_n =-(r_b+r_c)$ and
\begin{equation}\label{masslambda}
M = {r_br_c(r_b+r_c)\over 2 (r_b^2+r_c^2+r_br_c)},\qquad 
\Lambda = {3\over (r_b^2+r_c^2+r_br_c)}
\end{equation}
The function $f(r)$ is positive in the static region $r_b < r < r_c$,  and negative in the
cosmological and black hole regions, $r>r_c $ and $r<r_b$ respectively. 

Unlike the asymptotically flat or AdS cases,
when $\Lambda$ is positive there is a bound on the mass in terms of $\Lambda$ in order that two horizons exist. 
This is the limit in which the black hole and cosmological horizons approach each other, referred to as the Narai limit. Setting  $r_b =r_c $ in
equations (\ref{masslambda}) gives
\be\label{narai}
 M_N = { 1\over 3 \sqrt{\Lambda} } \ \  ,\quad r_N = 1/\sqrt{\Lambda} = 3M
 \ee
So in order to have a black hole horizon it is necessary that
\be\label{maxm}
3\sqrt{\Lambda} M\leq 1
\ee

The areas of the horizons are $A_h = 4\pi r_h^2$, $h=b,c$.
The horizon surface gravities $\kappa_h = f' (r_h ) /2 $ are given by
\begin{equation}\label{sg}
\kappa_b=  {(r_c-r_b)(2r_b+r_c)\over   2 r_b  (r_b^2+r_c^2+r_br_c)} \  \  ,\qquad
%T_b={\Lambda\over 12\pi r_b}(r_c-r_b)(2r_b+r_c),\qquad
\kappa_c= -  {(r_c-r_b)(2r_c+r_b)\over  2 r_c (r_b^2+r_c^2+r_br_c)}
%T_c={\Lambda\over 12\pi r_c}(r_c-r_b)(2r_c+r_b)
\end{equation}
Note that $ (r_b , r_c )$ and $(M , \Lambda )$ are two different parameterizations of the metric
If one wants to use $M$ and $\Lambda$ as the parameters, then the areas and surface gravities are implicit functions
of $( M, \Lambda )$ specified by the relation (\ref{masslambda})\footnote{One way of unravelling the relations is given in  \cite{McInerney:2015xwa}}

Note that $\kappa_b$ is positive and  $\kappa_c$ is negative. In black hole thermodynamics the horizon areas and surface gravities are
identified with the entropy and temperatures according to $S_h =A_h /4$ and $T_h = 2\pi \kappa_h$. Discussions of the meaning of 
negative temperatures in this context are given in, $e.g.$, \cite{Cvetic:2018dqf}\cite{Chakrabhavi:2023avi}.

\section{ Observing a black hole?}\label{redshiftstatic}
We start by reviewing the infinite redshift issue when trying to observe the region near the horizon of a black hole. Let a timelike observer ${\cal O}_I$, not necessarily geodesic, have four-velocity $u^a_I$ and let a photon have four-momentum $p^a$. Then the
energy of the photon as measured locally by observer ${\cal O}_I$ is given by
\be\label{energy}
E_I = - p^a u_I^b g_{ab}
\ee

 Consider two static observers in the metric (\ref{metric}), one located at $r_1$ near
the black hole and the other at $r_2$ farther away from the black hole. The four velocity of each is fixed by normalization to be
\be\label{staticfour}
u^t = {1\over \sqrt{f(r_I ) }} \ \  , \quad I=1,2
\ee
with other components zero.
Observer $\obsone$ emits a photon which propagates radially outward and is observed by 
$\obstwo$. The photon propagates along a null geodesic, and using the results of the next section (\ref{radpho2}), the four-momentum
of the photon is given by
\be\label{radpho}
p^t = {\nu \over f} \ \  , \quad p^r = \nu
\ee
where $\nu$ is a constant on the photon's path.
Then the ratio of the energies of the photon measured by the two observers, which is the inverse ratio of the wavelengths,
 is given by (\ref{energy}) and (\ref{radpho})
\be\label{redshift}
{E_{2 } \over E_{1 } }  ={\lambda_{1} \over \lambda_{2 } }={\sqrt{ f(r_1 )} \over \sqrt{ f(r_2 )} }
\ee
For $r_1$ near the black hole, $f(r_1 ) \rightarrow 0$.  In asymptotically flat spacetime, $f(r_2 ) \rightarrow 1$ for $r_2 \gg r_b$, and the ratio goes to zero.
Hence the light emitted from near the black hole is infinitely redshifted and becomes unobservable. In SdS, if $r_2$ is some location between the two horizons,
then $f(r_2 )$ is some finite value and the redshift is still infinite. 

If $\obsone$ is on the surface of a star that is collapsing to a black hole, although $\obsone$ is not static, the ratio of the energy of a photon emitted 
from the surface to the energy of the photon when received by a distant $\obstwo$, still goes to zero as the surface of the star falls within 
its own horizon. So at some point in the collapse process, light from the star and received by $\obstwo$ is redshifted below the detection limit, 
and the star appears forever suspended in a late stage of collapse. From the point of view of distant observers, the star never forms a black 
hole, and is often referred to as a ``frozen star". Of course $\obsone$ on the surface of the star measures it to pass through its horizon in finite
proper time.   

Returning to equation (\ref{redshift}), one notices that if instead of being a generic observer, if
 $\obstwo$ is close to the cosmological horizon, then $f(r_2 )$ is also going to zero.
This raises the possibility that the ratio could be non-zero. Explicitly, near each horizon $r_h$,
\be\label{fnear}
f(r) \simeq 2\kappa_h (r- r_h )
\ee
and the ratio of the energies for two static observers who are each near a horizon becomes
\be\label{redshiftnear}
{E_{2 } \over E_{1 } } \simeq \left( {\kappa_b (r_1  -r_b ) \over |\kappa_c |  (r_c - r_2 )  } \right)^{1/2}
\ee 
So by tuning the $|r-r_h |$ the observers can tune the redshift. 
The dependence on $M$ and $\Lambda$ is contained in the $\kappa_h$ and $r_h$.

Static observers are accelerated and so some force must be available to provide the acceleration. The ratio of the redshifts can be expressed
in terms of the acceleration of each observer, which is a locally measurable quantity. For a static observer the acceleration is radial,
$a^r = u^\nu \nabla_\nu u^r$, with $u^\nu$ given in (\ref{staticfour}). 
In an orthonormal frame this gives
\be\label{accel}
a^{\hat{r}} = {1\over 2} {f^\prime (r) \over f } 
\simeq   { \sqrt{ | \kappa_h | } \over  \sqrt{ 2 |r-r_h | } } 
\ee
where (\ref{fnear}) has been used in the last step. Hence the ratio of redshifts can be expressed in terms of the physical acceleration 
needed to keep each observer static,
\be\label{redshiftneartwo}
{E_{2 } \over E_{1 } } \simeq  \left( {\kappa_b \over | \kappa_c | }\right)^{3/2} {|a_2 | \over |a_1 |}
\ee
When this ratio is order one, we will say that 
it constitutes a ``good observation" by $\obstwo$ of the region near the black hole horizon. For very small black holes $\kappa_b \gg |\kappa_c | $, so
for the ratio to be order one needs $|a_2 | \ll |a_1 | $.  In the opposite limit when the black hole and cosmological horizons approach each
other, the surface gravities approach the common value of zero.  Then the ratio of the redshifts can be order one with each observer having 
the same magnitude accelerations, which could be maintained, $e.g.$, b a sturdy rope connecting the two 
observers.

\section{Geodesic motion}
Readers uninterested in the  derivations about geodesics can skip to Section 5, or simply look at results illustrated in Figures 1-3.

Static observers are accelerating. Geodesic observers are
more natural, or at least less expensive, not needing a force to produce the motion. 
 We proceed to analyze geodesic motion,
and then in Section 3 compute the redshifts between different pairs of geodesic observers. We will see that qualitatively the redshift/blueshift 
 effects discussed in the previous section still hold

Most of our analysis will be for positive $\Lambda$, but at certain points it is interesting to compare the results for asymptotically de Sitter, flat, and anti-de Sitter
spacetimes ($\Lambda$ positive, zero, or negative respectively). So we start with $\Lambda$ general.
Let $u^a$ be the tangent to a geodesic, with components $u^\alpha = (u^t , u^r , 0, u^\phi )$ in the coordinates of equation (\ref{metric}).

 The metric (\ref{metric}) has the two Killing vectors
\be\label{kvs}
\vec{\xi } ={\partial \over \partial t } \ \  , \quad  and \quad \vec{L} ={\partial \over \partial \phi } 
\ee
The inner product of a Killing vector and $u^a$ is constant on the path, so let
\be\label{constant}
\xi^a u^b g_{ab} = -C \ \  , \quad and\quad L^a u^b g_{ab} = \lt
\ee
where $\lt$ and $C$ are constants of the motion. $\lt$ is interpreted as the  angular momentum  per unit mass (for timelike particles).
 In the static patch of SdS where $\xi^a$ is timelike $C$ is interpreted as the energy per unit mass.
Equations (\ref{constant}) imply
\be\label{cons}
u^t = {C\over f(r)} \ \  , \quad  and \quad u^\phi = {\lt \over r^2 }
\ee

Using these relations and normalization of four-velocity, $u^a u^b g_{ab} = - \kappa$, with $\kappa =0$ for null and $\kappa =1$ for timelike $u^a$ gives
%Let $L$ denote the angular momentum of a timelike particle. Then the potential energy of such a particle is given by
 \be\label{rdot} 
(u^r )^2  = C^2  - V(r) 
 \ee
 where the potential $V$ is
 \be\label{potential}
   V(r) = \left(\kappa +\frac{ \lt^2}{r^2}\right)f(r)
\ee

It is convenient  to change to  the dimensionless radial coordinate
\be\label{xdef}
x= {r\over M}
\ee
The black hole horizon is smallest  when $M\ll 1/\sqrt{|\Lambda |}$ in which case  $r_b \rightarrow 2M$, so  $x>2$ outside the black hole.
The smallest black hole has the largest cosmological 
horizon with  $r_c \rightarrow \sqrt{ 3}/\sqrt{\Lambda}$ and $x\gg 1$. 

In the timelike case the potential as a function of $x$ is
 \be\label{potentialx}
   V(x) = -{2\alpha \over x^3} +{\alpha \over x^2} -{2\over x} +1 -\gamma\alpha -\gamma x^2
\ee
where the two dimensionless parameters $\gamma$ and $\alpha$ are
\be\label{dimpar}
\gamma =  \L3 M^2 \ \  ,\quad \alpha ={\lt^2 \over M^2 }
\ee
The possible forms of
 $V$ depend on the parameters. As $x\to 0$, $V\cong -{2\alpha \over x^3} \to -\infty $ for nonzero $\alpha$. As $x\to \infty$,
  $V\cong -\gamma x^2 \to \mp \infty$ for $\Lambda$ positive or negative respectively. When $\Lambda=0$
then $V\to 1$ at infinity. So the large distance behavior of the potential is markedly different depending on $\Lambda$, illustrating the cosmological 
expansion or the bounce back behavior.

\subsection{Radial geodesics}
When $\lt =0$ the timelike geodesics are given by
\be\label{radur}
u^t ={C\over f(r)} \ \  , \quad  u^r = \pm \sqrt{ C^2 - f(r) }   \quad , \quad \quad timelike
\ee
where the $\pm$ signs are for radially outward and inward respectively.  
The potential vanishes  for null radial geodesics  when $\lt =0$, so $ \dot{ r}^2  = C^2 $ and
\be\label{radpho2}
u^t = {C \over f} \ \  , \quad u^r = \pm C  \quad , \quad  \quad null
\ee

\subsection{Timelike Circular Geodesics}
Unstable and stable circular orbits occur at the maxima and minima of $V$ respectively, which depend on $\alpha $ and $\gamma$,
and bound orbits exist near a stable circular one.  The value of  $C^2$ in a circular orbit is determined by  $C^2 = V(r_0 )$. 
So the task at hand is to find the extrema of $V$.

 In the timelike case the derivative of $V$ is 
 \begin{align}\label{pdef}
{dV(x) \over dx}  & = {2\over x^4 } P(x) \\ \nonumber
%P(x) & = - \L3 M^2 x^5 + x^2 - {\lt^2 \over M^2 } x + 3{\lt^2 \over M^2 }
P(x) & = - \gamma x^5 + x^2 - \alpha x + 3\alpha
\end{align}
 Orbits are located at the positive  zeros of the fifth order polynomial equation $P(x) =0$.

\subsubsection{Timelike Static Geodesic}
All SdS spacetimes have
a static timelike geodesic, which can be thought of as the limit of a circular orbit when the angular momentum goes to zero. 
 Setting $\lt = \ \alpha =0$ in (\ref{pdef}) gives
\be\label{pstez}
P(x) = x^2 (-\gamma x^3 + 1) =\   0
\ee
with solution $x^3 = \gamma^{-1}$. There are three solutions in the complex plane, which can be chosen to be $( \gamma ^{-1/3},
\  \gamma ^{-1/3} e^{\pm 2\pi i /3} )$
when $\gamma>0$ and $ ( - |\gamma | ^{-1/3},\  |\gamma |^{-1/3} e^{\pm \pi i /3}) $ when $\gamma <0$. There is a real, positive
solution only when 
$\gamma  > 0$, 
\be\label{staticroot}
x_{st} = \gamma^{-1/3} =\    \left( {3\over \Lambda M^2 }\right)^{1/3}
\ee
This solution represents the balance between the inward gravitational attraction of the black hole and the
outward cosmological flow due to $\Lambda$. From a Newtonian point of view, the two forces are encoded in the geodesic equation, and the static solution to the 
geodesic equation corresponds to the forces balancing at an extremum of the potential. Here $x_{st}$ a maximum of the potential and so
is an unstable point. Related solutions in the extended SdS spacetime have been studied
 \cite{Faruk:2023uzs} in the context of understanding the instaton
contribution to the Euclidean path integral.

The timescale  of the instability is set by the negative frequency squared
\be\label{staticfrequ}
\nu^2 = {1\over 2} V''(r_{st} ) = -\Lambda
\ee
This is independent of the mass $M$, so when $M=0$  this reduces to the familiar result that $r=0$ is a static geodesic. 
References \cite{Pretorius:2007jn}\cite{Cardoso:2008bp}\cite{Cardoso:2003sw} identify this quantity as the Lyapunov exponent 
for the (in)stability of the geodesic. Note that in (\ref{staticfrequ}) the timescale is 
defined with respect to proper  time, while in these references the timescale is defined with respect to the Killing time $t$, which is appropriate 
for asymptotically flat spacetimes. That introduces a factor of $( dt/d\tau  )^2$ in the denominator of(\ref{staticfrequ}).
In SdS, $t$ is not a good coordinate on the cosmological horizon and becomes spacelike outside the horizon, so
 proper time is a natural alternative.

\subsubsection{Circular orbits: roots of $P$}

Stable and unstable circular orbits outside the black hole occur at the minima and maxima of the potential respectively and
bounded orbits exist near the minima. So we want to identify  the positive zeros of $P$. We proceed  with $\Lambda >0$, except in the perturbative
analysis below which holds for either sign of $\Lambda$.

At small and large argument $P(x)$ goes like
%
%\be\label{pasymp}
%P  \sim  3\alpha -  \alpha x  \ \  ,\quad x\rightarrow 0
%\ee
 \begin{align}\label{pasymp}
%P & \sim  3{\lt^2 \over M^2 } -  {\lt^2 \over M^2 } x  \ \  ,\quad x\rightarrow 0 \\
%P & \sim - \L3 M^2 x^5    \ \  ,\quad x\rightarrow \pm \infty
P & \sim  3\alpha -  \alpha x  \ \  ,\quad x\rightarrow 0 \\
P & \sim - \gamma x^5    \ \  ,\quad x\rightarrow \pm \infty
 \end{align}
 The real polynomial $P(x)$ has five zeros in the complex plane which occur in complex conjugate pairs, so there are either one, three, or five real roots, and we are
interested in real roots with $x>2$ (outside the black hole).
We know that in Schwarzschild there are two circular orbits when $\alpha > 12$, so one expects that those orbits will still exist for sufficiently small $ \gamma $. 
Hence a third zero of $P$ must also be present for small $ \gamma $.  Below we argue for
the existence of a root, for all $\gamma$, based on
the existence of the static geodesic (\ref{staticroot}). So these  arguments lead us to 
expect either one or three zeros of $P$. By considering the merging of roots
we then argue that there are only one or three roots, rather than five.

The number of real roots of the polynomial changes when two roots merge and then move off the real axis into the complex plane.
 When two roots coincide both $P=0$ and
  $dP/dx =0$ at a critical point $x_{cr}$. An example is given in Figure (\ref{fig:Lplots}), in which $\gamma$ is fixed and 
  the number  of zeros changes as $\alpha $ is varied.

The  two equations $P=0$ and
  $dP/dx =0$ can be solved for $\alpha$ and $\gamma$, giving
 
 \be\label{pzero}
  \gamma ( x_{cr}) = { x_{cr}-6 \over  x_{cr}^3 ( 4x_{cr} -15 ) } \ \  , \quad \quad \alpha ( x_{cr} ) = {3x_{cr}^2 \over   4x_{cr} -15 }
  \ee
The pair of equations  (\ref{pzero}) are a parametric representation for a curve of critical points in the $\gamma - \alpha$ plane which is 
plotted in Figure (\ref{fig:gamma,alpha}).

Since $\gamma$ and $\alpha$ must be positive we see that physically relevant solutions must have $x_{cr}  \geq 6$. 
At $x_{cr}  = 6$,   $\gamma =0$ and
$\alpha =12$, which correctly gives the case of Schwarzschild. The function $\gamma (x_{cr} )$ also goes to zero for $x_{cr}  \gg 6$ ,
so it has a maximum at some $x_{cr}  > 6$. Setting the derivative $\gamma ' (x_{cr} ) =0$ gives the quadratic equation
$2x_{cr}^2 -21 x_{cr} +45=0$. This has two roots, $x_{cr} $ equal to $3$ or $15/2$. So the physical root is $ x_{cr} =15/2$, which implies
that the maximum value of $\gamma$ at a critical point is
\begin{align}\label{maxgamma}
Max (\gamma (x_{cr}  )) & = \gamma (15/2 )  \\
& = 6.4 \cdot 10^{-3} Max (\gamma_{bh} ) \simeq 2.4 \cdot 10^{-4}
\end{align}
where $Max (\gamma_{bh}  ) =1/27 \simeq 3.7 \cdot 10^{-2}$ in the black hole parameter space. Hence the maximum value of $\gamma$ for which
there are three orbits is a small fraction of the possible range of $\gamma$, and in this sense, most of the $\gamma - \alpha$ phase space 
has only one, unstable, orbit. 
A plot of the curve of critical points (\ref{pzero}) is shown in Figure \ref{fig:gamma,alpha}. We know from graphing different examples of $P$, such
as in Figure \ref{fig:Lplots}, that there are choices of the parameters that give one or three zeros of $P$. Since the number of zeros can only
change on the critical curve (\ref{pzero}), the entire region under  the curve must have three  zeros, and the outside region must have one. 
\begin{figure*}
\includegraphics[width=\textwidth]{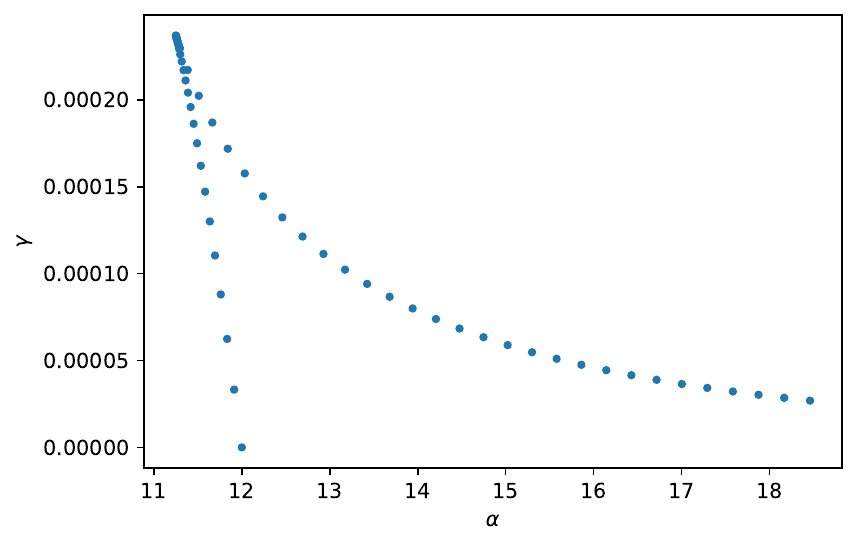} 
\caption{Curve of critical points showing $\gamma=\L3 M^2$ \ as a function of $\alpha= { \lt^2 \over M^2 }$. Under the curve there are 
three circular orbits (two unstable, one stable), on the curve there are two, and elsewhere 
 there is one. Recall that $\gamma < 3.7 \cdot 10^{-2} $ in the black hole parameter space. }
\label{fig:gamma,alpha}
\end{figure*}

To summarize,  in the portion of the parameter space
below the curve of critical points there are three orbits, on the curve there are
two orbits, and elsewhere in the parameter space there is one. Hence a stable orbit only occurs when there are three orbits. 
To see what happens in a fixed spacetime, $i.e.$, with fixed values of $M$ and $\Lambda$, look at a horizontal line $\gamma ={\Lambda \over 3} M = constant$.
  For example, choosing $\gamma = 1.2889 x 10^{-4}$ as for the plots in Figure (\ref{fig:Lplots}) one sees that 
at small angular momentum there is one orbit, at intermediate values of angular momentum there are three orbits, and at large values there is again 
one orbit. Each of these regions is divided by a critical point at which there are two orbits, corresponding to the intersections of the $\gamma =constant$
line with the curve of critical points.
 For a  value of $\gamma$ that is above the peak in Figure \ref{fig:gamma,alpha}, there is only one one 
 zero of $P(x)$ and so only one, unstable, circular orbit. 
\begin{figure*}
\includegraphics[width=\textwidth]{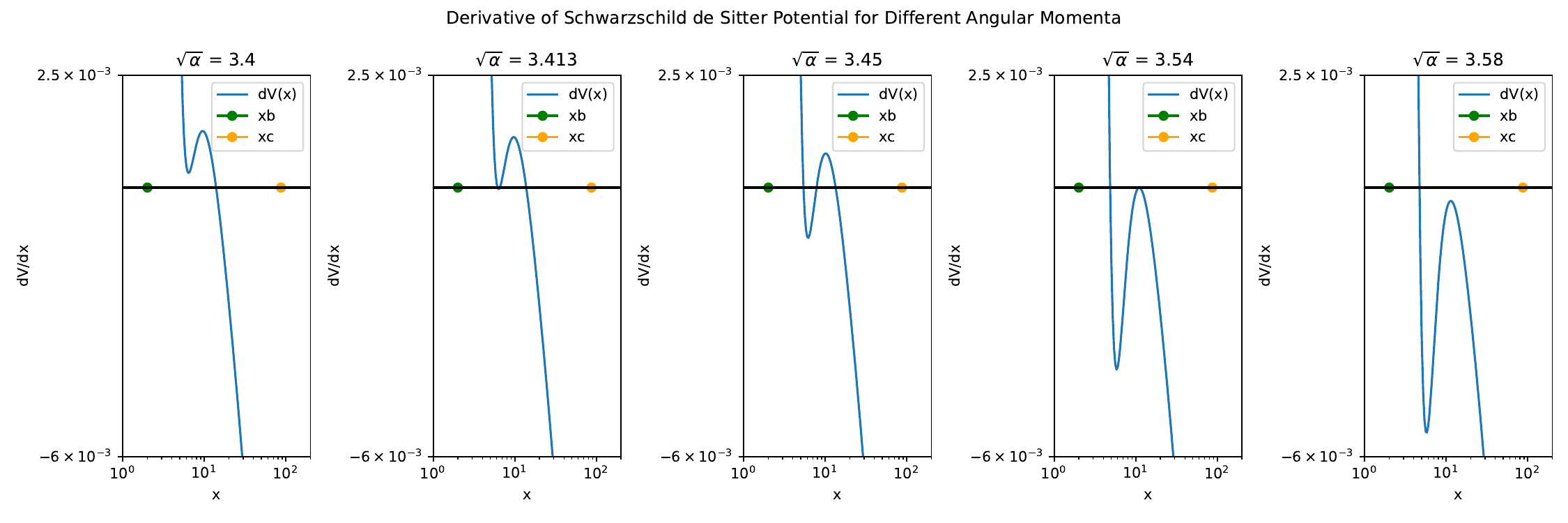} 
\caption{ Plots of $dV/dx =2P/x^4 $  as a function of distance $x=r/M$ for fixed $\gamma = 1.2889 \cdot10^{-4}$ are shown for
representative values of $\alpha = { \lt^2 \over M^2 }$. From the left, $\sqrt{\alpha}$ = 3.54, 
$\sqrt{\alpha}$ = 3.413, $\sqrt{\alpha}$ = 3.45, $\sqrt{\alpha}$ = 3.54, and $\sqrt{\alpha}$ = 3.58. Circular orbits occur where the curve crosses the 
$x$ axis.
The black hole horizon is marked by a green dot and 
the cosmological horizon is marked by an orange dot.  }
\label{fig:Lplots}
\end{figure*}

Fixing $M$ and varying $\Lambda $ corresponds to a vertical line of constant $\alpha$ in Figure \ref{fig:gamma,alpha},
and shows how the number of orbits changes with
$\Lambda$  at fixed angular momentum. For example, it is of interest to see how nonzero $\Lambda$ affects the situation compared to Schwarzschild.
Schwarzschild has two orbits for any  $\alpha > 12$, so consider an $\alpha_0 >12$ and fix $M$.  As $\Lambda $ increases from zero there are
three orbits, which decreases to two when $\alpha_0$ intersects the curve of critical points, and then there is only one orbit at larger $\Lambda$.
As $\gamma$ approaches its maximal value of $1/27$, the
 radius of this one orbit approaches the radius of the  static geodesic (\ref{staticroot}), which is also the limit of the radii of the black hole and 
 cosmological horizons, all converging to $r=3M$. 

The minimum value of $\alpha$ on the curve of critical points occurs at the maximum value of $\gamma$ on the curve,
\be\label{minalpha}
Min(\alpha (x_{cr}  )) = \alpha (15/2 )  = 11.25
\ee 
Again fixing $M$ and following a vertical line of constant $\alpha$ with $11.25 < \alpha < 12$, for small $\Lambda$ there is one orbit,
then three orbits inside the cusp of Figure \ref{fig:gamma,alpha}, then one orbit at larger $\Lambda$. For $\alpha < 11.25$ there is  
one orbit for all allowed $\Lambda$. This is in distinction to Schwarzschild where necessarily $\alpha >12$ for orbits, and reflects the fact that 
the cosmological expansion acts as an outward force which aids the angular momentum to balance the inward pull of the black hole.

\begin{figure*}
\includegraphics[width=\textwidth]{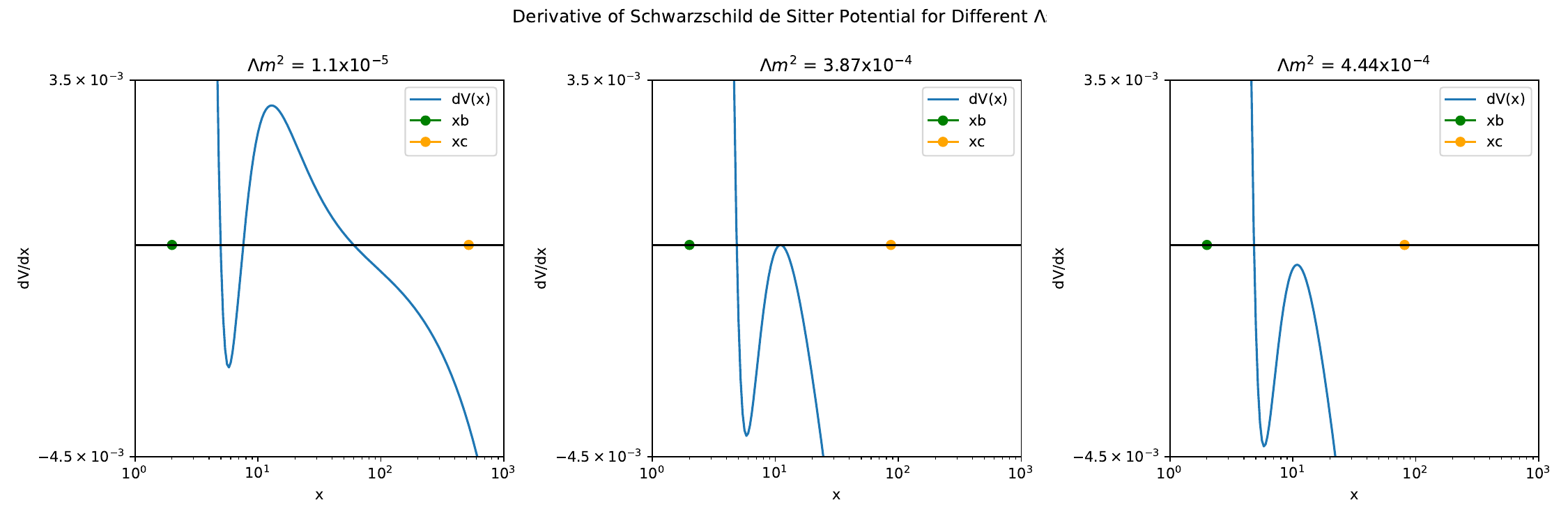} 
\caption{The derivative of the potential $ dV/dx =2P/x^4$ as a function of dimensionless distance $x$ for fixed $\alpha$ of 12.53 is shown for
different values of $\gamma =\L3 M^2$,
 $\gamma = 1.1\cdot 10^{-7}$ (left), $\gamma =  1.1\cdot 10^{-6}$ (middle) and $\gamma = 3.9\cdot 10^{-6}$ (right). The middle plot corresponds to the critical value of $\gamma$. Circular orbits occur where the curve crosses the 
$x$ axis.
 The black hole horizon is marked by a green dot  and the cosmological horizon is marked by an orange dot.  }
\label{fig:dv,f}
\end{figure*}

\subsubsection{Perturbative solutions for roots:}
\underline{Near Schwarzschild orbits}

The potential is the sum of the Schwarzschild term plus the $\Lambda$ dependent term, $i.e.$, $V=V_S + \Delta V$ with 
$\Delta V = -\gamma ( x^2 + \alpha )$.
%$\Delta V = -{\Lambda \over 3} M^2 ( x^2 + \lt^2 /M^2 )$.
In Schwarzschild spacetime with $\Lambda =0$,   $P=0$ reduces to the quadratic equation $ (x- x_{s+}) (x- x_{s-})=0$ with roots
\begin{equation}\label{schroots}
 x_{s\pm} =  {\alpha\over 2  }\left(1\pm \sqrt{1-{12\over\alpha }  }\right) \ \  ,  \quad   \Lambda =0
\end{equation}
%\begin{equation}\label{schroots}
% x_{s\pm} =   \frac{\lt^2}{2M^2} \left(1\pm \sqrt{1-\frac{12M^2}{\lt^2}}\right) \ \  ,\quad \Lambda =0
%\end{equation}
The solutions are real for $\alpha  \geq 12$, and merge to a single solution at equality.  So we expect  for small $\gamma $ and $\alpha > 12$
there will be two zeros close to the Schwarzschild ones, plus a third zero representing physics that is distinct from
Schwarzschild. We find  these solutions using perturbation theory.

We have
\be\label{Vtwo}
{dV \over dx}  ={2\over x^4} \left( (x- x_{s+ }) (x- x_{s-} ) -\gamma x \right) 
\ee
%\be\label{Vtwo}
%{dV \over dx}  ={2\over x^4} \left( (x- x_{s+ }) (x- x_{s-} ) -{\Lambda \over 3}M^2 x \right) 
%\ee
where $ x_{s\pm}$ are given in (\ref{schroots}).  Let 
\be\label{xpert}
x_{0\pm}=x_{s\pm}( 1 + \delta_\pm  )\ \  , \quad |\delta_{\pm} | \ll 1
\ee
Substituting $x_0$ into the equation $dV/dx = 0$ and keeping  terms  linear in $\delta$ gives
\be\label{pertroot}
\delta_\pm =\  \pm \gamma {x_{s\pm}^4 \over x_{s+}- x_{s-} } =\  \pm 96 \gamma \left( \alpha \over 12  \right)^3
 { \left[1\pm (1-12/ \alpha^2 )^{1/2} \right]^4 \over  (1-12/ \alpha^2 )^{1/2} }  \  , \quad \gamma \ll 1
\ee
%\be\label{pertroot}
%\delta_\pm =\  \pm \L3 M^2 {x_{s\pm}^4 \over x_{s+}- x_{s-} } =\  \pm 32\Lambda M^2  \left( \lt^2 \over 12 M^2 \right)^3
% {\left[1\pm (1-12M^2/ \lt^2 )^{1/2} \right]^4 \over  (1-12M^2/ \lt^2 )^{1/2} } 
%\ee
The approximate solutions (\ref{xpert}), (\ref{pertroot}) for circular orbits is valid for either sign of $\Lambda$ ($i.e.$, either 
sign of $\gamma$). Since $\delta_{\pm}$ is proportional to $\Lambda$ the
direction of the effect of a small $\Lambda$ is opposite in $dS$ compared to $AdS. $ In 
order that $\delta$ is small compared to one in (\ref{pertroot}), one needs that $\alpha /12 = \lt^2 / 12 M^2  $ is
sufficiently greater than one so that the denominator is not close to zero. This means that in the numerator the prefactor $(\alpha /12 )^3 $ is large, so
 $ \gamma = {\Lambda \over 3} M^2$ must be quite small so that the correction $\delta$ is small. This is consistent with Figures \ref{fig:Lplots} and
 \ref{fig:dv,f}  where it is found that the value of 
 $\gamma $ is very small in order that three roots can be discerned. 

\underline{Near static orbits}

As noted above there are either three or one circular orbits, so there must be a third zero of $P$ to go along with the two just found.
This will be an unstable orbit, and will be due to physics that is qualitatively different from the $\Lambda =0$ spacetimes. 
The static $\lt =0$ geodesic (\ref{staticroot}) gives a hint, since it is
a strictly SdS phenomenon, and it exists at any point on the sphere of radius $x_{st}$. This suggests that an orbit exists at a radius close to  $x_{st}$ 
 and with very small angular momentum, corresponding to moving the particle slowly on the static sphere. We look for such a solution
 perturbatively, assuming that $\alpha $ is small in (\ref{pdef}). Let
 \be\label{xperttwo}
 x_{st,L} = x_{st} (1 + \epsilon ) \ \  , \quad \quad x_{st} = \left( {3\over \Lambda M^2 }\right)^{1/3}
 \ee
 where $| \epsilon |   \ll 1$.
Substituting into $P(x) =0$, using the zeroth order relations (\ref{pstez}) and (\ref{staticroot}), and keeping terms linear in $\epsilon$ gives
\be\label{xstl}
\epsilon = -{ \lt^2 \over 3 M^2 }\left( {\Lambda M^2 \over 3} \right)^{1/3}  \left( 1- 3\left( {\Lambda M^2 \over 3} \right)^{1/3} \right)
\ee
Since $P$ is linear in $ \lt^2 \over 3 M^2$, the result (\ref{xstl}) is actually good as long as the prefactor is small,
\be\label{xstlcond}
{1\over 3} \alpha \gamma^{1/3}  =\  { \lt^2 \over 3 M^2 }\left( {\Lambda M^2 \over 3} \right)^{1/3} \ll 1
\ee

The fact that $\epsilon$ is negative makes sense in the basic force balance picture.  
Since $x_{st}$ results from balancing the inward and outward forces on a zero angular momentum mass, and since angular momentum gives
another outward ``force", the inward pull of the black hole must be increased to maintain the balance. 

This perturbative solution does not depend on whether $\Lambda$ is positive or negative.
In the AdS case the real $x_{st}$ root is negative, and this is perturbatively related to a real, negative root with nonzero $L$ in a similar manner.
Hence there are again three real roots of $P$ in the near Schwarzschild regime, just one of them is not relevant for a physical geodesic.

\subsection{Null Geodesics}
The motion of null geodesics is governed by the conservation equations (\ref{cons}) plus the equation for the radial motion (\ref{rdot}) with $\kappa =0$ in the potential,
\be\label{phopot}
V= \left( 1- {2M\over r} \right) {\lt ^2 \over r^2 }  -{\Lambda \over 3} \lt^2 
\ee
For radial geodesics  $\lt =0$, so $V=0$  and the solution for the four-momentum was  given in (\ref{radpho}). 

For $\lt$ nonzero,
the $\Lambda$ term  only contributes a constant to the potential, and so the derivative $V'(r)$ is independent of $\Lambda$ \cite{Islam:1983}
\footnote{ See \cite{Gibbons:2008ru}  for an analysis of the observational implications.}.
Therefore as in Schwarzschild, there is an unstable circular orbit at $r_0= 3M$, or $x_{0,null} =3$,
\be\label{nullorbit}
V' ( r_{0,null})  =0 \ \ \quad r_{0,null} =3M
\ee
The time scale of the instability, or Lyapunov exponent   \cite{Pretorius:2007jn}\cite{Cardoso:2008bp} for this orbit is set by 
\be\label{nulltime}
\nu^2 = {1\over 2} V''(r_{0,null} ) = {\lt^2 \over (3M)^4 }
\ee
Comparing calculations of quasinormal mode frequencies to the properties of the unstable null geodesic orbit  
 in asymptotically flat spacetime \cite{Cardoso:2008bp},  it is
shown that $\nu^2$ is proportional to the imaginary part of the QNM frequencies.
Subsequent calculations  of QNM in SdS in the near maximal mass limit
\cite{Cardoso:2003sw} provide evidence that this connection also holds in SdS.

As $\Lambda$ approaches 
the maximal value (\ref{narai}), $r_c$ and $r_b$ both approach $r_{0,null} =3M$ from above and below respectively and the three null surfaces converge.
In this sense, photons in the null orbit 
are probing the geometry closer to the black hole horizon than when $\Lambda =0$.

\section{Redshift for geodesic observers in the static patch}
In Section (\ref{redshiftstatic}) we computed redshift of a photon exchanged between two static observers, giving the results (\ref{redshiftnear}),
(\ref{redshiftneartwo}). Since a force is required to keep static observers in place, 
 geodesic observers are simpler -- or at least cheaper --than the static ones. References \cite{Wang:2023fge} and \cite{Cao:2024kht}  numerically
 integrate null geodesics to construct  images and plots of what would be observed at various locations $r_2$. 
 These are quite interesting from the point of view of our calculations.  The authors note that there is an
 enhancement in luminosity of the received signals as  $r_2$ gets closer to the horizon, and \cite{Cao:2024kht} relates this to the redshift enhancement
 factor in the intensity.
 
 So let us again consider a photon propagating radially outwards, 
 emitted by $\obsone$ at $r_1$  and received by  $\obstwo$ at $r_2$, with $r_2 > r_1$.
Let the two observers be in geodesic motion. The four momentum of the photon is given in  (\ref{radpho}) and the
four velocity of the observers in (\ref{cons}) and (\ref{rdot}). Substituting into (\ref{energy}) for the energy of the photon measured locally  by each observer
gives
 \be\label{redshiftgen}
{E_{2 } \over E_{1 } } = 
{C_2\over C_1 }\cdot { f(r_1 ) \over  f(r_2 ) }\cdot
 {1 - u^r (r_2 ) /C_2^2  \over   1 - u^r (r_1 ) /C_1^2 } \quad , \quad (radial \ photon)
 \ee
where $u^r  (r)= \pm \sqrt{1 - V(r ) /C^2 }$. In a circular orbit, as well as for the static geodesic  (\ref{staticroot}), $u^r =0$. The potential $V(r)  $ depends on
$\lt$ and is given in (\ref{potential}), and $C_I$ and $\lt_I$ 
are constants of the motion for observer $I$. Unless otherwise stated, we assume the observers are timelike.
We proceed with the analysis of redshifts measured by various choices of geodesic observers, all located in the static patch of the spacetime.

\subsection{Radial Motion}\label{radialmotion}
 Let $\obsone$ $\obstwo$ be in radial motion, with $\obsone$ falling into the black hole. Using (\ref{radur}) in  (\ref{redshiftgen})
 the ratio of observed energies becomes
\be\label{redshift2}
{E_{2 } \over E_{1 } } = 
{ f(r_1 ) \over  f(r_2 ) }\cdot
 {C_2 \mp \sqrt{C_2^2 - f(r_2 )  }
 \over  C_1 + \sqrt{ C_1^2 - f(r_1 )  } } \ \  ,\quad radial \  geodesic\  observers
\ee
where the $\mp$ option is minus if $\obstwo$ is falling outwards and plus for inwards, and necessarily $C_I^2 > f(r_I ) $ at the the position  $r_I$
of the observer. In particular, we want to consider
 $r_1$  close to $r_b$ and $r_2$ close to $r_c$, so that $f(r_I )  \simeq 2\kappa_h (r - r _I  ) $ which is going to zero.
Then also assuming that $C_I^2 \gg |2\kappa_h (r - r _I  )|$, we have
\be\label{redshift3}
{E_2 \over E_1} \simeq  {\kappa_b (r_1 - r_b ) \over \kappa_c (r_2 -r_c )} \cdot
{ C_2 \mp [ C_2 - | \kappa_c | (r_c -r_2 )/ C_2  ]\over 2 C_1 - \kappa_b (r_1 -r_b )/ C_1 }
\ee
If $\obstwo$
is moving outwards (the minus case) through the cosmological horizon the factor of $(r_2 - r_c )$  in the denominator is cancelled by the
one in the numeration, and the ratio is proportional to $(r_1 - r_b )$, which 
goes to zero as $\obsone $ gets closer to the black hole. That is, the redshift goes to infinity and the black hole becomes unobservable.

However if $\obstwo$ is falling inwards moving towards the photon  (the plus case), then the ratio becomes
 \be\label{redshift4}
{E_2 \over E_1}  \simeq  {C_2 \over C_1} {\kappa_b (r_1 - r_b ) \over \kappa_c (r_2 -r_c )}  
 \ee
Now the zero in the numerator can  be cancelled by the zero in the denominator to give a finite ratio, $i.e.$, signals from near the black hole can be detected
by using the blue shift of the cosmological horizon. Equation (\ref{redshift4}) is one of our main results. This is similar in spirit to the result for two
static observers (\ref{redshiftnear}), though the dependence on the $\kappa_I (r_I -r_h) $ is linear for the freely falling observers compared to 
a square root dependence for the accelerated observers. In (\ref{redshift4}) one has the additional freedom of tuning $C_2 / C_1 $.

Comments on the range of
 validity of the near horizon approximations are given in the Appendix.

\subsection{Observers in circular or static orbits}
 The orbits that are
closest to the two horizons are both unstable, but in the interest of getting the best measurements, let's 
assume that the receiver $\obstwo$ can maintain position in a large circular orbit for long enough to make measurements.
Suppose that emission from $\obsone$ happens via particles in an accretion disc slipping to lower energy orbits in the process emitting photons 
at frequencies known in the particle rest frame. Then using (\ref{redshiftgen}) with $u^r =0$ the ratio of the observed photon energies is
\be\label{circred}
{E_2 \over E_1 } = {C_2 \over C_1} {f(r_1 ) \over f(r_2 ) } 
\ee
The smallest circular orbit allows the closest look at the black hole, and is at $r=3M$ for ultra-relativistic particles and photons.
The largest orbit yields the most blue shift. It's radius  approaches the static value of $r_{st}$ given in
equations (\ref{staticroot}) and (\ref{xperttwo}).  Having fixed the values of  the $r_I$ then fixes the value of the $C_I$ through the relation.
$(u^r )^2 = 0= C_I^2 -V(r_I )$, with $V$ given in (\ref{potential}). So $C_2^2 = V(r_2 ) \simeq f(r_{st} )$ and  $C_1^2 = V(3M ) \simeq (\lt^2 / 9M^2 ) f(3M)$ 
for highly relativistic particles. The latter approximation for $V(3M) $ requires that
\be\label{approx}
{9 M^2 \over \lt_1^2 } \ll 1
\ee
Substituting into (\ref{circred}) gives
 \begin{align}\label{circred2}
{E_2 \over E_1 }  & = {3M \over \lt_1 } { \sqrt{ f(3M) } \over \sqrt{ f( r_{st} ) } }\\  \nonumber
&= {\sqrt{3} M \over \lt_1 }  \left( 1 + 3 \gamma^{1/3} + 9 \gamma^{2/3} \right)\  ,\quad \quad \gamma={\Lambda M^2 \over 3}
\end{align}
The condition (\ref{approx}) implies that
the prefactor in (\ref{circred2}) is small. (If it is meaningful for $\obsone$ to be null, there is no such constraint.) 
The last factor, the polynomial in $ \gamma^{1/3}$, varies from one  to three as $\gamma$ varies from zero to $1/27$. 
\footnote{ We note a curious point. The ratio $f(r_1 ) / f(r_2 )$
is the ratio of two polynomials, and so in general is a rational function in $\gamma$. However for the smallest and largest orbits,
  $r_1 = 3M$ and $r_2 =r_{st} = M /\gamma^{1/3}$, 
$f(3M) $ factors and  the ratui
 rational function reduces to 
a  polynomial in $ \gamma^{1/3}$, giving the simple expression in (\ref{enhance}). }
Hence 
this polynomial is the enhancement factor due to positive $\Lambda$,
\be\label{enhance}
 \left({E_2 \over E_1 } \right)_{\Lambda >0 } = \left( 1 + 3 \gamma^{1/3} + 9 \gamma^{2/3} \right) \left({E_2 \over E_1 } \right)_{\Lambda =0}
\ee
The enhancement due to $\Lambda$ varies between one and three.

To summarize, while signals emitted from the smallest circular and received by observers in the largest circular orbit gain a modest enhancement from
a positive cosmological constant, the best strategy for observing emission from near a black hole is for the receiver to be close to the
cosmological horizon and falling inwards, towards the black hole.

\section{Redshift for geodesic observers in the cosmological patch}\label{cosmo}
In this section we show that if $\obstwo$ is in the cosmological patch, just outside the cosmological horizon and moving
radially inwards, the redshift works similar to the  situation in subsection (\ref{radialmotion}) when the observer is just inside the cosmological horizon. 
Hence the results of the previous section extend continuously to the region outside the  cosmological horizon.
In the cosmological patch radial geodesic motion can again be inwards or outwards, but now an observer who starts close to the horizon and falls
inwards still gets further away from the horizon,  which moves away from the observer at the speed of light. 

\subsection{geodesics in good coordinates}
To study null rays that cross the cosmological horizon one needs to use coordinates that are good in a neighborhood of the horizon.
In reference \cite{Gregory:2018ghc} a new time coordinate $T$ was introduced for SdS that is good on both horizons. Let
\begin{align}\label{newcoord}
T & = t + h(r ) \  , \quad  with\ \  h'(r ) = \eta / f \\
r' & =r
\end{align}
where $f$ is as before (equation (\ref{fdef})) and 
\begin{equation}\label{etadef}
 \eta (r) =  -\gamma  r +{\beta \over r^2} \  ,\quad\quad \gamma = { r_c ^2 + r_b ^2 \over r_c ^3 - r_b ^3}\  \   , \quad \beta = {r_c^2 r_b^2 (r_c  + r_b ) \over r_c ^3 - r_b ^3 }
\end{equation}
Since $r'=r$ we will drop the primes. 
 The metric in the new coordinates is
\begin{equation}\label{metrictwo}
ds^2 = -f dT^2 +  2\eta dr dT + {dr^2 \over f} ( 1-\eta^2  ) +r^2 d\Omega^2
\end{equation}
The constants $\gamma$ and $\beta$ were fixed by requiring that the metric is well behaved on the two horizons, which
turns out to be the conditions
\be
\eta (r_b ) =1 \  , \quad \eta (r_c ) = -1
\ee
It can be checked that $T$ interpolates between ingoing and outgoing Eddington-Finklestein coordinates between the two horizons.
The function $\eta$ satisfies  $\eta > 1$ for  $r<r_b $ and $\eta < -1$ for  $r>r_c $. 
The time translation Killing vector is
\be\label{KVT}
\xi^a = {\partial \over \partial t } =   {\partial \over \partial T }
\ee
The Killing vector $\xi^a $ is spacelike in the cosmological region,
so for  geodesic observer $\obstwo$ with four-velociy $u_{(2)}^a$ in the cosmological region we set
\be\label{constant2}
  \xi^a u_{(2)}^b g_{ab} = B_2
  \ee
while for $\obsone$ in the static patch we keep the definition 
$\xi^a  u_{(1)}^b g_{ab} = -C_1 $ in equation (\ref{constant}). Of course changing the name $C_2 =-B_2$ does not change the physics, but
our end formulas are nicer in that physical quantities that should be positive are positive when $B_2 >0$.
In the new coordinates equation (\ref{constant2}) gives for $\obstwo$
\be\label{uT}
u^T_2  = {1\over f } (  B_2 + u^r_2 \eta ) 
\ee
This relation is the same as the first of (\ref{cons}), using $dT = dt + \eta / f dr$. 
The equation for $dr / d\tau$ is the same as (\ref{rdot}) since that results from the two scalar relations
and depends only on $r$. One can check this directly by substituting (\ref{uT}) 
 into $u^a u^b g_{ab} = -\kappa $, so $u^r_2 $ is again given by (\ref{radur}).
 
 The components of the null geodesic in the coordinates (\ref{metrictwo})) are
   \be\label{photonnew}
   p^T   = {\nu \over f } ( 1 + \eta ) \ \  \ , \quad   p^r  =  \nu  \ \  ,\quad \nu = constant
\ee
so the observed energy of the photon, defined in (\ref{energy}),  is $E_2 = - ( \nu / f)(  B_2 + u^r )$.
 
  What is gained by switching to the new coordinates is that we can compute the redshift for photons that cross the cosmological horizon.
  So again consider a photon propagating radially outwards, emitted by $\obsone$ who is falling into the black hole. The photon is received by  $\obstwo$,
who is located in the cosmological patch. From our previous results we know that the best strategy for $\obstwo$ to get a small redshift, and so
a detectable measurement, is to
travel on a radial  geodesic going inwards, 
\be\label{cosmoin}
u^r_2 = - \sqrt{ B_2^2 - f(r_2 ) }  
\ee

Calculating the ratio of observed energies for $\obsone$ close to $r_b$ and $\obstwo$ close to $r_c$ gives the result analogous
to  (\ref{redshift4}),
 \be\label{redcosmo}
{E_2 \over E_1}  \simeq  {B_2 \over C_1} {\kappa_b (r_1 - r_b ) \over | \kappa_c  |(r_2 -r_c )}  
 \ee
 which is positive for $B_2$ and $C_1$ positive, and the fact that $\kappa_c$ is negative has been taken into account. Comparing to (\ref{redshift4})
we see that the redshift is the same if $\obstwo$ is just inside or just outside the cosmological horizon.

\section{Conclusion}
In contrast to
a baseball park, suitable for rock concerts,
some concert halls have particularly good acoustics and faithfully transmit sound from the orchestra to listeners.
There are even architectural oddities that contain audio focusing, like the famous ``whispering galley" in the U.S. Capitol. In an analogous way,
gravitational fields 
act as lenses for light, and in this paper we have studied the arrangement of nested black hole and cosmological horizons in 
affecting the locally measured frequency of photons. This geometry shows that in principle a star collapsing to a black hole can be observed by 
outside observers. The best results, $i.e.$, the most discernible measurements are made by an accelerated observer who stays close
 to the cosmological horizon. Good measurements can also be made by inward traveling geodesic observers while they are close to the cosmological horizon. 
We also found that signals received by observers in the largest available circular geodesic orbits are enhanced by the cosmological constant. These
orbits are unstable, so apparently collecting data about the region arbitrarily close to the black hole horizon is possible, but not easy.
 
 \section{Appendix}
In the case that the two geodesic  observers are in radial motion, their coordinate locations $r_1$ and $r_2$ are changing with time.
The near horizon approximation for $f$ (\ref{fnear}) has been used, 
but the validity of the approximation changes along the path.
Since $\obsone$ is heading into the black hole the near horizon approximation gets better along that path,
 but $\obstwo$
is moving away from the cosmological horizon, so at some $r_{2f}$ the ratio gets too small and $\obstwo$ can no longer make a good measurement.
 To estimate when that happens, we look at the second 
 term in a Taylor series expansion of $f$ in equation (\ref{fnear}), and see that the approximation requires 
 \be\label{taylor}
\lvert {f'' (r_c ) \over 2 f'(r_c ) } (r- r_c )  \rvert = {1\over r_c }\cdot { 1- {r_b \over r_c} -{r_b^2 \over 2 r_c^2 } \over 1+ {r_b \over r_c} +{r_b^2 \over 2 r_c^2 }  } 
|r-r_c | \ll1
\ee
where equations (\ref{masslambda}) have been used.
Checking this for the small $r_b /r_c \ll 1$ and large black hole $r_b /r_c \rightarrow 1$ limits, the condition (\ref{taylor}) reduces to
\be\label{cond}
 {|r_{2f} -r_c | \over r_c }\ll1
\ee
When $\obstwo$ reaches an $r_{2f}$ such that this ratio is order one, signals emitted from near the black hole have a large redshift.

The change in coordinate distances for the two observers corresponds to proper time intervals
governed by  (\ref{radur}). When $C_I^2 \gg 2 \kappa_h (r-r_h )$ this reduces to\footnote{One does not gain much by using the 
exact expression for the integral which in the near horizon
region is $ -{1\over | \kappa_I } [  \sqrt{C_I^2 - 2\kappa_h (r- r_h) }]_i ^f $}  
\be\label{rtau}
|\Delta r_I | \simeq C_I \Delta \tau_I \ \ ,  \quad I=1,2
\ee
and so
 \be\label{redshift5}
{E_2 \over E_1}   \simeq  {\kappa_b \over  | \kappa_c |}  {\Delta \tau_2 \over \Delta \tau_1 } 
 \ee
 The ratio of the surface gravities $\kappa_b / | \kappa_c |$ ranges from very large for small black holes to order one for large black holes. 
To summarize, equations (\ref{redshift4}) and (\ref{redshift5}) imply that a static patch observer $\obstwo$ can get a picture of the near black hole region,
by using the blueshift of the cosmological horizon.

%\bibliographystyle{alpha}
%\bibliography{Sources.bib}

\begin{thebibliography}{99}


%\cite{Hackmann:2010zz}
\bibitem{Hackmann:2010zz}
E.~Hackmann, C.~Lammerzahl, V.~Kagramanova and J.~Kunz,
``Analytical solution of the geodesic equation in Kerr-(anti) de Sitter space-times,''
Phys. Rev. D \textbf{81}, 044020 (2010)
doi:10.1103/PhysRevD.81.044020
[arXiv:1009.6117 [gr-qc]].
%\cite{Lammerzahl:2015qps}
\bibitem{Lammerzahl:2015qps}
C.~L\"ammerzahl and E.~Hackmann,
``Analytical Solutions for Geodesic Equation in Black Hole Spacetimes,''
Springer Proc. Phys. \textbf{170}, 43-51 (2016)
doi:10.1007/978-3-319-20046-0\_5
[arXiv:1506.01572 [gr-qc]].


%\cite{Momennia:2023lau}
\bibitem{Momennia:2023lau}
M.~Momennia, A.~Herrera-Aguilar and U.~Nucamendi,
``Kerr black hole in de Sitter spacetime and observational redshift: Toward a new method to measure the Hubble constant,''
Phys. Rev. D \textbf{107}, no.10, 104041 (2023)
doi:10.1103/PhysRevD.107.104041
[arXiv:2302.11547 [gr-qc]].


%KdS
% The authors study images seen by observers in circular orbits
 %of KdS BHs derived from numerically integrating the geodesic equation. They note that the intensity of the image increases
% with increasing $\gamma$ (at fixed observer position), and as the observer moves closer to the cosmological horizon ( for fixed $\gamma$).
%\cite{Wang:2023fge}
\bibitem{Wang:2023fge}
K.~Wang, C.~J.~Feng and T.~Wang,
``Image of Kerr{\textendash}de Sitter black holes illuminated by equatorial thin accretion disks,''
Eur. Phys. J. C \textbf{84}, no.5, 457 (2024)
doi:10.1140/epjc/s10052-024-12825-3
[arXiv:2309.16944 [gr-qc]].




%SdS and RNdS
% The authors study images seen by observers in circular orbits
 %of SdS and RNdA BHs derived from numerically integrating the geodesic equation. They note that the intensity of the image increases
% as the observer moves closer to the cosmological horizon ( for fixed $\gamma$).
%\cite{Cao:2024kht}
\bibitem{Cao:2024kht}
L.~M.~Cao, L.~Y.~Li, X.~Y.~Liu and Y.~S.~Zhou,
``Appearance of de Sitter black holes and strong cosmic censorship,''
Phys. Rev. D \textbf{109}, no.8, 084021 (2024)
doi:10.1103/PhysRevD.109.084021
[arXiv:2401.15408 [gr-qc]].

%\cite{Faruk:2023uzs}
\bibitem{Faruk:2023uzs}
M.~M.~Faruk, E.~Morvan and J.~P.~van der Schaar,
``Static sphere observers and geodesics in Schwarzschild-de Sitter spacetime,''
JCAP \textbf{05}, 118 (2024)
doi:10.1088/1475-7516/2024/05/118
[arXiv:2312.06878 [gr-qc]].

%\cite{McInerney:2015xwa}
\bibitem{McInerney:2015xwa}
J.~McInerney, G.~Satishchandran and J.~Traschen,
``Cosmography of KNdS Black Holes and Isentropic Phase Transitions,''
Class. Quant. Grav. \textbf{33}, no.10, 105007 (2016)
doi:10.1088/0264-9381/33/10/105007
[arXiv:1509.02343 [hep-th]].

%\cite{Cvetic:2018dqf}
\bibitem{Cvetic:2018dqf}
M.~Cveti{\v{c}}, G.~W.~Gibbons, H.~L{\"u} and C.~N.~Pope,
``Killing Horizons: Negative Temperatures and Entropy Super-Additivity,''
Phys. Rev. D \textbf{98}, no.10, 106015 (2018)
doi:10.1103/PhysRevD.98.106015
[arXiv:1806.11134 [hep-th]].

%\cite{Chakrabhavi:2023avi}
\bibitem{Chakrabhavi:2023avi}
V.~Chakrabhavi, M.~Etheredge, Y.~Qiu and J.~Traschen,
``Constrained spin systems and KNdS black holes,''
JHEP \textbf{02}, 231 (2024)
doi:10.1007/JHEP02(2024)231
[arXiv:2311.07777 [hep-th]].

%shows that with a positive cosmological constant stable circular orbits only exist in a bounded region of the $\M, \Lambda , L$ parameter space.
%\cite{Howes:1979}
\bibitem{Howes:1979}
R. ~J.~ Howes, 
“Existence and stability of circular orbits in a Schwarzschild field with
non-vanishing cosmological constant”,
Australian Journal of Physics \textbf{32} (1979).

%\cite{Pretorius:2007jn}
\bibitem{Pretorius:2007jn}
F.~Pretorius and D.~Khurana,
``Black hole mergers and unstable circular orbits,''
Class. Quant. Grav. \textbf{24}, S83-S108 (2007)
doi:10.1088/0264-9381/24/12/S07
[arXiv:gr-qc/0702084 [gr-qc]].

%\cite{Cardoso:2008bp}
\bibitem{Cardoso:2008bp}
V.~Cardoso, A.~S.~Miranda, E.~Berti, H.~Witek and V.~T.~Zanchin,
``Geodesic stability, Lyapunov exponents and quasinormal modes,''
Phys. Rev. D \textbf{79}, no.6, 064016 (2009)
doi:10.1103/PhysRevD.79.064016
[arXiv:0812.1806 [hep-th]].

%\cite{Cardoso:2003sw}
\bibitem{Cardoso:2003sw}
V.~Cardoso and J.~P.~S.~Lemos,
``Quasinormal modes of the near extremal Schwarzschild-de Sitter black hole,''
Phys. Rev. D \textbf{67}, 084020 (2003)
doi:10.1103/PhysRevD.67.084020
[arXiv:gr-qc/0301078 [gr-qc]].

%\cite{Islam:1983}
\bibitem{Islam:1983}
J. ~N. ~Islam, “The cosmological constant and classical tests of general relativity”, Phys. Lett. \textbf{97A} (6) 239
(1983)

%\cite{Gibbons:2008ru}
\bibitem{Gibbons:2008ru}
G.~W.~Gibbons, C.~M.~Warnick and M.~C.~Werner,
``Light-bending in Schwarzschild-de-Sitter: Projective geometry of the optical metric,''
Class. Quant. Grav. \textbf{25}, 245009 (2008)
doi:10.1088/0264-9381/25/24/245009
[arXiv:0808.3074 [gr-qc]].

%\cite{Gregory:2018ghc}
\bibitem{Gregory:2018ghc}
R.~Gregory, D.~Kastor and J.~Traschen,
``Evolving Black Holes in Inflation,''
Class. Quant. Grav. \textbf{35}, no.15, 155008 (2018)
doi:10.1088/1361-6382/aacec2
[arXiv:1804.03462 [hep-th]].

\end{thebibliography}

\end{document}